\documentclass{article}
\usepackage{a4}
\usepackage{latexsym}
\usepackage{amsmath}
\usepackage{amsfonts}
\usepackage{amssymb}
\newcommand{\be}{\begin{eqnarray}}
\newcommand{\ee}{\end{eqnarray}}
\newcommand{\nn}{\nonumber \\}
\newcommand{\til}{\tilde{~}}
\newcommand{\di}{\mbox{dim}}
\newcommand{\diag}{\mbox{diag}}
\newcommand{\openone}{\mbox{1\kern -0.25em l}}
\newcommand{\openZ}{\mathbb{Z}}
\newcommand{\openR}{\mathbb{R}}
\newcommand{\openH}{\mathbb{H}}
\newcommand{\openC}{\mathbb{C}}
\newcommand{\CL}{C\!\ell}
\begin{document}
\title{
On the equivalence of Daviau's space Clifford algebraic, 
Hestenes' and Parra's formulations of (real) Dirac theory}
\author{Bertfried Fauser\\
Universit\"at Konstanz\\
Fakult\"at f\"ur Physik\\
Fach M 678\\
78457 Konstanz {\bf Germany}\\
Electronic mail: Bertfried.Fauser@uni-konstanz.de
}
%\date{preliminary version \today}
\date{August 31, 1999}
\maketitle
\begin{abstract}
Recently Daviau showed the equivalence of ordinary matrix based
Dirac theory --formulated within a spinor bundle 
${\cal S}_x \simeq \openC^4_x$--, to a Clifford algebraic
formulation within space Clifford algebra $\CL(\openR^3,\delta)
\simeq M_2(\openC) \simeq {\cal P} \simeq $ Pauli algebra
(matrices) $\simeq \openH \oplus \openH \simeq$ biquaternions.
We will show, that Daviau's map $\theta : \openC^4 \mapsto 
M_2(\openC)$ is an isomorphism. Furthermore it is shown that 
Hestenes' and Parra's formulations are equivalent to Daviau's space
Clifford algebra formulation, which however uses outer automorphisms.
The connection between such different formulations is quite 
remarkable, since it connects the left and right action on the
Pauli algebra itself viewed as a bi-module with the left 
(resp. right) action of the enveloping algebra ${\cal P}^e 
\simeq {\cal P} \otimes {\cal P}^T$ on ${\cal P}$. The isomorphism
established in this article and given by Daviau's map does clearly
show that right and left actions are of similar type. This should be
compared with attempts of Hestenes, Daviau and others to interprete the
right action as the iso-spin freedom. 
\end{abstract}

\noindent{\bf MSCS:} 15A66


\section{\protect\label{SEC-1}Introduction}

There is a long quest on a geometric intuitive description of
Dirac spinor fields. Only a few month after the 
publication of Dirac's first paper \cite{Dirac-1} Charles Galton
Darwin tried to re-express the {\it strange new objects} called
{\it half vectors} by Pauli \cite{Pauli-hdb} and spinors due to
Paul Ehrenfest --according to B.L. van der Waerden, see
\cite{BudinichTrautmann}-- with help of tensors \cite{Darwin}. He did not
fully succeed in obtaining an equivalence by writing down complex tensor
equations which yield Dirac's theory ``twice over'' --with a doubling of
degrees of freedom from complexification--; see Parra
\cite{Parra-dirac} for a detailed review on this topic.
Madelung, trying the same transcription essentially reproduced
Darwin's results, most likely without knowing them
\cite{Madelung}. Also in the thirties Fock and Ivanenko 
\cite{Fock} did very important work on the geometric relations
behind the $\gamma$--algebra introduced by Dirac. De~Broglie and his 
school developed a very valuable and complete picture of the Dirac fluid
--a tensor description of the Dirac field-- and its hydrodynamics
\cite{Yvon,Takabayasi}. This reasoning has a revival in recent times
because of the improved tool of Clifford algebra now available
\cite{Rylov}.

But the historical development abandoned the attempt of trying
to find a geometric --and thereby tensorial-- description of the
Dirac field. Firstly there seemed to be a tendency to concrete
calculations which on the one hand were extremely successful
and on the other hand could be performed without an elaborated
interpretation by applying simply the {\it rules} of
$\gamma$--algebra, see discussion in \cite{Isham}. Furthermore, 
quantum theory {\it had (has?) to be\/} interpreted within a statistical
picture. It was simply out of the imagination of that time to search for
such an explanation or even to connect geometry with spinor variables. 

One has to wonder, but neither the physicists Pauli and Dirac
nor the mathematicians Weyl, Jordan, von Neumann and others cited
or seemed to have known substantially the work of Grassmann,
Clifford, Klein, Cayley, Hamilton and other algebraists of the 19th
century. If some of their formulas and results were
acknowledged --the quaternions e.g. were well known to be
isomorphic to Pauli matrices-- this was done in a technical
sense. The geometric origin of hypercomplex number systems was unknown
or ignored and thus lost for a further development of the
theory. One result of this missed opportunity --in the sense of
Dyson \cite{Dyson}-- was the thereby obtained ``interpretation''
of spinors, which became artificial objects in an {\it abstract
spin space\/} or an {\it inner spin space\/} and had thusly 
no physical counterpart in the ``real world''. However, from
a technical point of view, dropping the interpretation, there was
an extraordinary and fruitful development of spinor methods in
physics.

The situation changes with the appearance of the writings
of David Hestenes \cite{Hest-1}, see references in \cite{Hest-n}. 
He recovered again the geometric origin of spinor objects and the 
formerly well known connection of (metric) space and certain algebras. 
The first time he gave a geometrical motivated treatment of real
Dirac theory in his book ``Space time algebra'' \cite{Hest-1}. The
reformulation of Dirac's theory in real(!) space time algebra
$\CL(\openR^4,\eta)$, $\eta=\diag(1,-1,-1,-1)$ is the starting
point of a host of new insights into the interplay between
geometry, algebra and physics. Hestenes' reformulation was also
the very first starting point of Daviau's consideration which
lead to a space algebraic ($\simeq \CL(\openR^3,\delta)$),
$\delta=\diag(1,1,1)$ formulation of Dirac theory.

But even up to now, there is a discussion on the proper
interpretation of spinorial objects in either geometrical or
statistical settings. This lead to a large number of slightly
different notations of spinors; e.g. spinor modules $S_x \simeq
\openC^{2n}_x$, operator or Hestenes spinors $\simeq \CL^+_{p,q}$,
ideal spinors $\simeq \CL f$, $f$ an primitive idempotent
element, algebraic spinors and the spin Clifford bundle
--isomorphism classes of ideal spinors to geometrically
equivalent idempotents-- etc. If Clifford algebra provides us
the {\it universal language for mathematics and physics}\/
\cite{Hest-ulmp} we have to give exact and unambiguous notations
of physical objects and of their exact mathematical design.

Hestenes in succeeding to write down a {\it real} Dirac theory
within $\CL_{1,3}$ translated the non-geometrical $i=\surd{-1}$
into the {\it right action} of $\gamma_2\gamma_1$ --recall
$(\gamma_2\gamma_1)^2= -\gamma_1^2\gamma_2^2=-1$--.  But right
actions mix different left ideals related to different
idempotents, while left action remains in the same left ideal.
Rodrigues et al. introduced therefore the spin Clifford bundle
and algebraic spinors, in which spinors or even better algebraic
spinors are defined to be equivalence classes of ideals which 
belong to geometrically equivalent idempotents
\cite{Rodrigues}. Such idempotents are conjugated to one
another within the Clifford--Lipschitz group $\Gamma$ by
$e^\prime=ueu\til$, $u\in \Gamma$,$\til$ the reversion map, 
and are therefore members of the same group orbit. To obtain a
mathematical clear picture one should then translate the Dirac--Hestenes
spinors into the quotient space $DH \simeq \CL_{1,3}/\Gamma$ (as linear
space) to be not troubled with the probably ill chosen representants. This
consideration should, however, be compared with the approach of
Parra to Dirac--Hestenes spinors and his illuminating
explanation of the equivalence classes and their relations to
the Wigner definition of a {\it particle} as an irreducible
representation of the Poincar\'e group \cite{Parra-dirac}.  

In this paper, we want to study the map from ordinary Dirac matrix theory
onto space Clifford algebra used by Daviau. This will be done in several 
steps. Starting with the definition of the Daviau map, we analyse
afterwards the Hestenes formulation of Dirac theory. Then it is shown,
that a special option of Parra's formulation corresponds directly to
Hestenes' formulation, showing the well known correspondence between them.
Finally the equivalence of Hestenes', sic. Parra's, formulation to the
space Clifford algebraic formulation of Daviau is demonstrated. The
correct identification to Parra's options is given.

We can however not appreciate every work concerned with space Clifford
algebraic formulations of Dirac theory for lack of space, one importand
paper may be added here for those \cite{Baylis1997}.

Our analysis unmasks a close connection between the ordinary spinor modul 
$S_x\simeq \openC^4_x$ which is equivalent to a formulation by ideal
spinors in $\CL_{4,1}$, since $\CL_{4,1} \cong {\bf M}_4(\openC)$ which
is actually used by physicists. Daviau's map furthermore shows up a
correspondence of left actions on $\openC^4_x$ spinors to homomorphisms of
${\cal P}$, which can be written as $uxv$, $u,v,x \in {\cal P}$. If one 
defines the enveloping algebra ${\cal P}^e$ as in Hahn \cite{Hahn},
${\cal P}^e\cong {\cal P}\otimes {\cal P}^T$, where ${\cal P}^T$ 
denotes the right module or transposed module, it is easily seen, 
that the ${\cal P}^e$ left action is equivalent to the 
${\cal P}$-bi-module structure by writing 
${\cal P}^e\bullet{\cal P} \mapsto {\cal P}$, 
$x\otimes y^T\bullet z = xyz$. We have therefore to consider left and 
right actions on ${\cal P}$, as Daviau did. This makes ${\cal P}$ a 
${\cal P}$-bi-module. This bi-module structure is crucial for further 
investigations of the enveloping algebra ${\cal P}^e$ of Clifford
algebras, which will be given elsewhere, and for a thoughtful
interpretation of left and right actions in Clifford algebras. There is a
widespread thinking about the meaning of right actions, see 
\cite{Hestenes-iso,Daviau,Fauser-iso}.

\section{\protect\label{SEC-2} The Daviau map $ \theta : \openC^4
\mapsto \CL_{3,0}$}

\subsection{Definition of the Daviau map}

Daviau changed his notation and got rid of his cyclic permuted
$\sigma$-matrices in a new work \cite{Daviau1998}, however, we 
stay with his old notations to be coherent.

We start according to Daviau with the Dirac equation in its
standard matrix representation due to Bjorken \& Drell
\cite{BjorkenDrell} 
\be
&-i\gamma^\mu\partial_\mu \Psi + qA^\mu\gamma_\mu \Psi +m\Psi=0.&
\ee
We have $m,q$ real constants, $i=\surd{-1}$ the usual complex
unit, $\partial_\mu:=\partial/\partial x^{\mu}$ the partial
derivatives with respect to a local holonom coordinate system,
$A^\mu$ real components of an external vector potential,
$\Psi$ is the Dirac spinor of $\openC^4$ valued functions of the
(tangent) Minkowski space and finally $\gamma_\mu$ the Dirac
matrices in Dirac representation
\be\label{PM}
&\gamma_0=\gamma^0:=
\left(\begin{array}{cc} \openone & 0 \\ 0 & -\openone 
                \end{array}\right) 
\hskip 0.5cm
\gamma_k=-\gamma^k:=
\left(\begin{array}{cc} 0 & -\sigma^k\\ \sigma^k & 0  
                \end{array}\right) 
\hskip 0.5cm
\openone:=\openone_{2\times 2}&\nn
&\sigma^1:=\left(\begin{array}{cc} 0 & 1\\ 1 & 0  
                \end{array}\right);
\hskip 0.5cm
\sigma^2:=\left(\begin{array}{cc} 0 & -i \\ i & 0 
                \end{array}\right);
\hskip 0.5cm
\sigma^3:=\left(\begin{array}{cc} 1 & 0 \\ 0 & -1 
                \end{array}\right).&
\ee
It is an easy task, to translate the Dirac equation into a set of
eight {\it real} coupled differential equations, see also
\cite{Parra-conf}. From a mathematical point of view, this two
sets of equations are identical. But in setting
\be\label{RS}
&\Psi=\left(\begin{array}{c} 
 \Psi_1 \\ \Psi_2 \\ \Psi_3 \\ \Psi_4 
             \end{array}\right) 
:=      \left(\begin{array}{c}
a+ie \\ -g-if \\ d+ih \\ b+ic 
             \end{array}\right)&
\ee
with $a,\ldots,h : (M,\eta) \mapsto \openR$ real valued
functions, one does no longer insist on the ``spinorial''
character of the object in favour for playing with components
and forgetting about transformation properties -- compare the
analysis of Parra \cite{Parra-dirac,Parra-conf}.

On the other hand, one has to consider the Pauli algebra or
space Clifford algebra $\CL_{3,0}\simeq {\cal P}$. This algebra
is isomorphic to the full matrix algebra $M_2(\openC)$ and thus
eight dimensional over the reals. A purely dimensional
comparison yields ${\di}_{\openR} \,\Psi = 8 = \di \, {\cal P} =
{\di}_{\openR} \, M_2(\openC)$. 

The aim of the Daviau map is to give an isomorphism from $\openC^4_x
\rightarrow$ co-ordinates $\rightarrow {\bf M}_2(\openC)$ which is also a
morphism of the {\it algebraic\/} structure. One could call such a map a
Dirac-morphism.

Now, by letting
\be\label{phiD}
&u:=a+ih,\hskip 0.5cm v:=f+ib, \hskip 0.5cm 
 w:=c+ig,\hskip 0.5cm t:=d+ie&\nn
&\phi_1:=u+w,\hskip 0.5cm \phi_2:=t+v,\hskip 0.5cm
 \phi_3:=t-v,\hskip 0.5cm \phi_4:=u-w&\nn
&\phi_D=\left(\begin{array}{cc}
\phi_1 & \phi_3 \\ \phi_2 & \phi_4 \end{array}\right) \in
M_2(\openC) \simeq {\cal P} \simeq \CL_{3,0},&
\ee
we obtain a map $\theta : \openC^4 \mapsto M_2(\openC)$.
Introducing then (note our indexing)
\be\label{def-vec}
&\nabla:=\partial_0+\vec{\partial},\hskip 0.5cm
\vec{\partial}:=
\sigma_2\partial_1+\sigma_3\partial_2+\sigma_1\partial_3&\nn 
&A:=A_0+\vec{A},\hskip 0.5cm
\vec{A}:=
A^1\sigma_2+A^2\sigma_3+A^3\sigma_1&\nn
&\phi^*:=\left(\begin{array}{cc}
\bar{\phi_4} & -\bar{\phi_2} \\ -\bar{\phi_3} & \bar{\phi_1}
\end{array}\right) = \sigma_2 \bar{\phi} \sigma_2&\nn
&i:=\sigma_1 \sigma_2 \sigma_3,\hskip 0.5cm [i,X]_-=0
\quad\forall X \in {\cal P},&
\ee
we obtain the space Clifford or Pauli algebraic form of Dirac's
equation due to Daviau:
\be
\nabla\phi i\sigma_1 = m\phi^* + qA\phi.
\ee
Daviau showed, that all transformation properties and
requirements are fulfilled within this picture, making his map finally a
Dirac-morphism preserving the algebraic structure of Dirac theory. A  
Lagrangian formulation is also possible. Using the above given
representation of Pauli matrices (\ref{PM}) one can reconstruct
an algebraic expression of the $M_2(\openC)$ matrix $\phi_D$.
From (\ref{phiD}) we find
\be\label{phi-alg}
\phi_D&=&\left(\begin{array}{cc}
         u+w & t-v \\  t+v & u-w \end{array}\right)\nn
       &=&\left(\begin{array}{cc}
         a+c+i(h+g) & d-f+i(e-b) \\  
         d+f+i(e+b) & a-c+i(h-g) \end{array}\right)\nn
       &=&a\openone+d\sigma_1+b\sigma_2+c\sigma_3
         +ei\sigma_1-fi\sigma_2+gi\sigma_3+hi.
\ee
This form of the Daviau spinor will be used below to show the equivalence
to other formulations.

\subsection{Hestenes equation}

We may further notice, that since $\di\, \CL_{1,3}=16$ and
$\di\, \CL^+_{1,3}=8$, $\CL^+_{1,3}$ may also be used as a
target for a map $ H : \openC^4 \mapsto \CL^+_{1,3}$. This
algebra $\CL^+$, called even subalgebra, consist of Dirac--Hestenes 
operator spinors and has in a natural manner a bimodul structure 
under the action of even elements. With the above choice of names 
for the real spinor components (\ref{RS}) we obtain the correspondence 
using $\gamma_{ij}:=\gamma_i\gamma_j$, $\Sigma_i:=\gamma_i\gamma_0$,
$i:=\Sigma_1\Sigma_2\Sigma_3=\gamma_{0123}$:  
\be\label{DSP}
\Psi_H&=&
 a + b\gamma_{10} +c\gamma_{20} + d\gamma_{30}
 +e\gamma_{21} + f\gamma_{23} + g\gamma_{13} + h\gamma_{0123}.\nn
&=&
 a+b\Sigma_1+c\Sigma_2+d\Sigma_3
 -fi\Sigma_1+gi\Sigma_2+ei\Sigma_3+hi
\ee
Where we have used the identities
\be\label{SIG}
&i\Sigma_1=i\gamma_{10}=-\gamma_{23},\hskip 0.5cm
 i\Sigma_2=i\gamma_{20}= \gamma_{13},\hskip 0.5cm
 i\Sigma_3=i\gamma_{30}= \gamma_{21}&
\ee
and anticipated the names of the variables in an appropriate manner to
fit into the Daviau scheme. The translated Dirac equation reads 
($m=m_0c/\hbar, q=e/\hbar c, \partial=\gamma^\mu\partial_\mu,
A=\gamma^\mu A_\mu$) 
\be\label{DHeq}
&\partial \Psi_H\gamma_{21}= m\Psi_H \gamma_0 + qA \Psi_H,&
\ee
which is the famous Dirac--Hestenes equation and representation
free. The elements on the right hand side of $\Psi_H$ describe
the spin bivector $S:=\gamma_{21}$ and the ``particles'' (local) 
velocity $v:=\gamma_0$ --a time-like vector measuring proper-time-- 
and do {\it not} fix a representation. For a discussion of the
relation between quantum logic, measurement and the choice of a
time-like direction in Dirac theory see \cite{Haft,Saller}.

Now, we may left multiply (\ref{DHeq}) by $-\gamma_0$ which
turns the equation (beside the mass term) into the space part of
the algebra. Using (\ref{SIG}) and
\be
&-\gamma_0\partial=-\gamma^0\gamma^\mu\partial_\mu
                 =\Sigma_\mu\partial_\mu&\nn
&-\gamma_0 A     =-\gamma^0\gamma^\mu A_\mu
                 =\Sigma_\mu A_\mu&
\ee
we remain with
\be\label{DHeq2}
&\Sigma_\mu\partial_\mu\Psi_H i\Sigma_3 =
- m \gamma_0 \Psi_H \gamma_0 +q\Sigma_\mu A_\mu\Psi_H&\nn
&\Sigma_\mu\partial_\mu\Psi_H i\Sigma_3 =
- m \Psi_H^\dagger           +q\Sigma_\mu A_\mu\Psi_H,&
\ee
which is written now within the space sector only. The transformation
$\Psi_H^\dagger \,=\, \gamma_0 \Psi_H \gamma_0$ represents the
hermitian adjoint, which is not an inner automorphism of the
Pauli algebra isomorphic to $\CL^+_{1,3}$, as indicated by the odd
element $\gamma_0$. 

This form of the Dirac-Hestenes' formulation will be needed in the proof
of the isomorphy to Daviau's formulation below.

\subsection{Parra's analysis of Dirac theory}

Parra analyzed the Dirac equation also in terms of a real set of
eight differential equations \cite{Parra-conf}. Like Darwin and
Madelung he afterwards tried to reinterpret this set of
equations in terms of vector analysis, --spinors {\it versus\/}
multi-vectors \cite{Parra-dirac}--. The novelty of Parra's approach is,
that he succeeded in formulating tensorial equations without any
complexification and thereby no doubling of degrees of freedom.
This is achieved by a simple inspection of the resulting eight
real equations. Under the assumption, that the real part $\Re(\Psi_1)$ 
of $\Psi_1$ --first component of the $\openC^4_x$ Dirac spinor--
transforms as a {\it scalar\/} quantity, the full set of
eight equations admits a vectorial character. The result is at
first not satisfactory since some terms remain to be only third
components of vectors. By {\it introducing\/} the {\it spin vector\/} 
$\vec{n}=(0,0,\hbar)\,(=-i S)$, one obtains a full $SO(3)$ rotationally
invariant set of vector equations. Denoting the two scalar quantities as
$\alpha,\lambda$ and the two vectorial quantities as
$\vec{E}=(E_1,E_2,E_3), \vec{B}=(B_1,B_2,B_3)$ one arrives at the
Parra type $\{0\}$ spinor
\be
&\Psi_{\{0\}}=\left(\begin{array}{c}
\alpha+iB_3 \\ -B_2+i B_1 \\ E_3+i\lambda \\E_1+i E_2
\end{array}\right).&
\ee
Now, it is purely a matter of choice which type of vector component
--scalar, first, second or third vector component-- one asserts
for $\Re(\Psi_1)$. The other three possibilities yield by the
same procedure, also introducing the spin-vector $\vec{n}$,
equally well suited spinor--tensor translations. A suitable
{\it choice\/} of names for the involved scalars and vectors yields:
\be&
\begin{array}{ll}
\Psi_{\{0\}}=\left(\begin{array}{c}
\alpha+iB_3 \\ -B_2+i B_1 \\ E_3+i\lambda \\E_1+i E_2
\end{array}\right) & 
\Psi_{\{2\}}=\left(\begin{array}{c}
B_2+i B_1 \\ \alpha -iB_3 \\ E_1 - i E_2 \\ -E_3 +i \lambda 
\end{array}\right) 
\\
\Psi_{\{1\}}=\left(\begin{array}{c}
E_1-iE_2 \\ -E_3+i\lambda \\ B_2+iB_1 \\ \alpha-iB_2
\end{array}\right) &
\Psi_{\{3\}}=\left(\begin{array}{c}
E_3+i\lambda \\ E_1+iE_2 \\ \alpha+iB_3 \\ -B_2+iB_1
\end{array}\right) .
\end{array}
&\ee
If we now introduce a basis $\{e_i\}$ with Clifford algebraic
relations $e_ie_j+e_je_j=2\eta_{ij}$ and the above notations for
$m$ and $q$, one obtains four {\it different\/} equations:
\be&
\begin{array}{lrl}
\{2\} &
 \nabla\Psi_{\{2\}} e_{21} +q A \Psi_{\{2\}} +m \Psi_{\{2\}} e_0=0
& \quad e^+_\uparrow \\
\{0\} &
-\nabla\Psi_{\{0\}} e_{21} +q A \Psi_{\{0\}} +m \Psi_{\{0\}} e_0=0
& \quad e^+_\downarrow \\
\{3\} &
 \nabla\Psi_{\{3\}} e_{21} -q A \Psi_{\{3\}} +m \Psi_{\{3\}} e_0=0
& \quad e^-_\uparrow \\
\{1\} &
-\nabla\Psi_{\{1\}} e_{21} -q A \Psi_{\{1\}} +m \Psi_{\{1\}} e_0=0
& \quad e^-_\downarrow .
\end{array}
&\ee
In the second column we give the identification --due to Parra-- with 
``particles'' associated with the corresponding equations.
$\pm$ indicates electron or positron where $\uparrow\downarrow$
indicates spin up or down --this is a choice, one might exchange the
meanings. The second of these equations --Parra option $\{2\}$-- 
happens to be the Dirac-Hestenes equation (\ref{DHeq}) if we identify
the $\{e_i\}$ and $\{\gamma_\mu\}$ bases, which thereby includes the
spin explicitly. The other three equations are new. Even if they are
similar in structure one is not able to remove the relative changes in
sign if two or more of these equations are considered at the same time.
Once more, we see the right action of the spin-bivector $e_{21}$ and of
the velocity vector $e_0$. One should note, that proceeding from Dirac
theory to quantum electrodynamics (QED), it became necessary to
introduce particle and antiparticle creation and annihilation operators
for each spin polarisation. While in QED the formalism takes care of
the different types of spinors, a simple complex linear combination
--as quite common in Dirac matrix theory!-- intermingles the
different Parra options without any chance to re-obtain them as different
equations.

The Parra spinors can easily be put within a quaternion basis.
Let $1,i_k:=ie_k$ be a quaternion basis, then the spinors of
r-option become $\Psi_r=q^1_r + i \bar{q}^2_r$ where $\bar{~}$
means quaternion conjugation. Since Hestenes spinors are
elements of $\CL^+_{1,3} \subset \CL_{1,3} \simeq M_2(\openH)$,
this can be extended to matrix spinors
\be
&\Psi_{\{r\}} = \left(\begin{array}{cc}
q^1_r & -\bar{q}^2_r \\
\bar{q}^2_r & q^1_r
\end{array}\right).&
\ee
The $2\times 2$ matrix structure is a matrix representation of 
the complex structure $(1,i)$. 

Since the Hestenes equation is formulated within abstract
algebra and not within a representation it is trivially
representation independent. But a change of bases has to be not
only an algebra isomorphism but moreover a Clifford algebra
isomorphism. Only elements of the Clifford-Lipschitz group
$\Gamma_{1,3}$ induce such transformations. Denoting the group
of even such elements as $\Gamma^+_{1,3}$, we expect the
quotient $D=\Gamma_{1,3}/\Gamma_{1,3}^+$ to be exactly the {\it
discrete\/} group of transformations which connect the Parra
options. Such transformations are beside the identity space
inversion, charge conjugation and time reversal.

We would thus submit, that the spin Clifford bundle defined by
Rodrigues et al. \cite{Rodrigues} is a slightly to large structure, 
since it does not properly distinguish the different particle types of
Parra. The ``spin-particle'' Clifford bundle should consist of
equivalence classes of idempotents with respect to an {\it even}
geometrical equivalence relation. The commutator relation and
thus the Clifford structure can be seen to be invariant under
discrete --or more generally odd-- transformation of the
Clifford-Lipschitz group \cite{Crumeyrolle}. 

Since we have thus established the equivalence of Parra's
equations --and the spin Clifford bundle-- in essence to the
Hestenes formulation, we concentrate now on the connection of
Hestenes' and Daviau's space Clifford algebraic formulations. The
Daviau space Clifford algebra form of Dirac's equation will
correspond directly to Parra option $\{1\}$ as will be shown below. 

\subsection{Equivalence of space Clifford and Hestenes formulation}

We will calculate the action of the outer automorphism within
the even algebra.  Therefore we compare the $\gamma_0$ action
with the action of $*$ introduced in (\ref{def-vec}) on the
Daviau spinor (\ref{phiD}). Observe the relation:
\be
\phi_D^*&=&\sigma_2 \bar{\phi}_D \sigma_2\nn
        &=&\left(\begin{array}{cc}
          a-c+i(g-h) & -d-f+i(e+b) \\
          f-d+i(b-e) & a+c+i(-h-g) \end{array}\right)\nn
        &=&a\openone-d\sigma_1-b\sigma_2-c\sigma_3
          +ei\sigma_1-fi\sigma_2+gi\sigma_3-hi.
\ee
Now, let us use the injection $\sigma_i \mapsto \sigma_i \otimes
\openone$, which gives a $4\times 4$ representation of the space
Clifford algebra, we are able to introduce a $\gamma_0$ in this
representation, thereby identifying $\Sigma$ and $\sigma$ elements.
However, this is no longer an element of the space Clifford algebra. We
can calculate
\be
\gamma_0 \phi_D^* \gamma_0&=&a\openone+d\sigma_1+b\sigma_2+c\sigma_3
          +ei\sigma_1-fi\sigma_2+gi\sigma_3+hi\nn
          &=& \phi_D,
\ee
by comparing with (\ref{phiD}). This might be rewritten as
\be
\phi_D^* &=& \gamma_0 \phi_D \gamma_0
\ee
and used in the rewriting of the Dirac-Hestenes equation
(\ref{DHeq2}) which then yields the Pauli or space Clifford
algebraic equation
\be
\Sigma_\mu\partial_\mu \Psi_H i\Sigma_3 &=& -m \Psi_H^* 
+ q \Sigma_\mu A_\mu \Psi_H.
\ee
To obtain the full equivalence between this formulation of the
Dirac-Hestenes theory to the space Clifford algebraic version of
Daviau, we have to perform two further steps.

The first is to explain the additionally minus sign in front of
the mass term. Redefining the sign of charge and angular
momentum measurement, i.e. $e\mapsto -e$, $\hbar \mapsto
-\hbar$, results in the appropriate change. Of course, from a
particle point of view this two particles are {\it not}
identical. They have a relation as a spin up electron to a spin
down positron and do correspond to different types of Parra
options in rewriting Hestenes' theory \cite{Parra-conf}. Since
no weak interactions are involved here, one can physically not
distinguish these options and there is no harm in this settings.
However, one should note that Daviau got four different
equations within his calculations, and there may be the chance
that one of them fit exactly to Hestenes theory without changing
the sign of the mass term.

The second step is a relabelling of base elements in a cyclic
way. This can be done by defining
\be
z &:& \sigma \mapsto \Sigma \nn
&&z(\openone)=\openone\nn
&&z(\sigma_i)=\Sigma_{i-1}\quad \mbox{cyclic}.
\ee
The map $z$ can be extended as an outer-morphism, that is a grade
preserving extension \cite{HestSob}, to the whole algebra by
setting $z(\sigma_i \sigma_j)=z(\sigma_i)z(\sigma_j)$ etc. Since
$z$ is a cyclic permutation, we have $z^3=1$ and $z^{-1}=z^2$.
It is crucial to note, that even if in the definition of the $*$
morphism in (\ref{def-vec}) via complex conjugation followed by
a transformation with $\sigma_2$, $*$ is not inner, it commutes
with $z$. That is we have $z(\phi^*)=z(\phi)^*$.

We obtain the following isomorphism noticing from
(\ref{phi-alg}) and (\ref{DSP}) that $z^{-1}(\Psi_H)=\phi_D$ holds:
\be
\Sigma_\mu \partial_\mu \Psi_H i\Sigma_3 &=& 
-m \Psi_H^* + q \Sigma_\mu A_\mu\Psi_H
\ee
acting by $m\mapsto -m$ and $z^{-1}$ results in
\be
&(\sigma_0\partial_0
+\sigma_2\partial_1+\sigma_3\partial_2+\sigma_1\partial_3)
\phi_D i\sigma_1 \quad =&\nn
&m \phi_D^* + q(\sigma_0 A_0+\sigma_2 A_1+\sigma_3 A_2
+\sigma_1 A_3)\phi_D&
\ee
which results with (\ref{def-vec}) in
\be
\nabla \phi_D i\sigma_1 &=& m\phi_D^* + q A \phi_D.
\ee
This proves the equivalence of Daviau's space Clifford algebraic
and Hestenes' formulation of Dirac's theory.

\section{Related work}

There seems to be a notorious revival of the transition between spinor and
tensor descriptions of Dirac theory. As we mentioned Darwin and Madelung,
there are far more also recent such approaches of which we will mention
only two more. Based on ideas of Sallhofer \cite{Sallhofer}, Simulik et
al. \cite{Simulik} used extensively a spinor--tensor transition, called
there {\it Maxwell--Dirac isomorphism\/}, in applications and some theoretical
investigations. Since their formalism is a restriction of the approach
developed by Parra, however not so detailed and pedagogical, we have
nothing more to prove there. Our preference is however not intended
to provide any priority claims.

A detailed thoughtful description of geometric electron theory with many
citations and critical remarks can be found in Keller \cite{Keller}.

A further genuine and important approach to the spinor-tensor transition
was developed starting probably with Crawford by P. Lounesto,
\cite{Lounesto} and references there. He investigated the question, how
a spinor field can be reconstructed from known tensor densities. The major
characterization is derived, using Fierz-Kofink identities, from elements
called {\it Boomerangs\/} --because they are able to come back to the
spinorial picture. Lounesto's result is a characterization of spinors
based on multi-vector relations which unveils a new unknown type of spinor. 

However, we want to submit, that even the notion of a multi-vector is quite
questionable in Dirac theory \cite{Fauser-dirac} and in general
\cite{Fauser-hecke}. The $\openZ_n$-grading used to define multi-vectors is
{\it not\/} a feature of Clifford algebra. One expects very different
spinor structures if different $\openZ_n$-gradings are properly
implemented \cite{Fauser-abla,Fauser-vacua}.

\section{Conclusion}

In this paper we discussed the isomorphism between spinor and multi-vector
formulation of Dirac theory. We proved the equivalence of Daviau's space
Clifford algebraic and Hestenes' operator spinor formulations of Dirac
theory as their equivalence to different special options of Parra's 
treatment. The important observation is, that in usual formulations the 
spinor representations are made up from left actions, while Daviau's 
formulation requires the bi-module structure of left {\it and\/} right 
actions. A detailed mathematical analysis of this fact will be given 
elsewhere \cite{Fauser-envalg}. Regarding iso-spin, which was sometimes 
introduced as right action, our analysis shows that one should be very 
careful in doing so.

A further remarkable fact is that the Daviau spinor is of the most general
form --most general element in the algebra-- and utilizes the full Pauli
algebra as representation space. This should be compared with the Hestenes
even operator spinors and ideal or column spinors which span the
representation space but {\it not\/} the algebra itself. It is peculiar at
this point carefully to distinguish representations and abstract algebra.
In this sense, Daviau's formulation is the most compact formulation which
can be found.

However, we gave some references which critically discussed the concept of
multi-vectors or $\openZ_n$-gradings in Clifford algebras. One knows that
different $\openZ_n$-gradings can produce quite different spinor modules.
This fact renders the unquestioned multi-vector structure as a peculiar
one. A careful study of the representation theory and their dependence on
gradings in such cases is required.

\section*{acknowledgements}

This work was supported by the Deutsche Forschungsgemeinschaft DFG
providing a travel grant to Zacatecas and Ixtapa Conferences in Mexico,
June/July 1999. Suggestions provided by C. Daviau%and J.M. Parra 
are greatefully acknowledged.


\scriptsize
\begin{thebibliography}{99}
%
\bibitem[Baylis 1997]{Baylis1997}
E. Baylis; {\it Eigenspinors and electron spin\/} in proceedings ``The
theory of the electron'' Cuautitlan / Mexico 1995, J. Keller, 
Z. Oziewicz eds. Adv. in Appl. Clifford Alg. 7(Suppl.) 1997 197--213
%
\bibitem[Bjorken et al. 1964]{BjorkenDrell} 
J.D. Bjorken, S.D. Drell; {\it Relativistische Quantenmechanik\/}
Mc-Graw-Hill Inc. [1964] German: BI-Wissenschaftsverlag /Mannheim 1966
%
\bibitem[Budinich et al. 1988]{BudinichTrautmann}
P. Budinich, A. Trautmann; {\it The spinorial chessboard\/}
Triest Notes in Physics, Springer 1988
%
\bibitem[Crumeyrolle 1990]{Crumeyrolle}
A. Crumeyrolle; {\it Orthogonal and sympletic Clifford algebras -- spinor
structures\/} Kluwer /Dordrecht  1990
%
\bibitem[Darwin 1928]{Darwin}
C.G. Darwin; {\it The wave equations of the electron\/}
Proc. Royal Soc. London A118 1928 654
%
\bibitem[Daviau 1998a]{Daviau1998}
C. Daviau; {\it Application \`a la th\'eorie de la lumi\`ere de Louis de
Broglie d'une r\'e\'ecriture de l'\'equation de Dirac\/}
Ann. de la Fond. Louis de Broglie 23(3/4) 1998 121--127
%
\bibitem[Daviau 1998b]{Daviau}
C. Daviau; {\it Sur les tenseurs de la th\'eorie de Dirac en
alg\`ebre d'espace\/}
Ann. de la Fond. Louis de Broglie 23(1) 1998 27--37\\
{\it Sur l'\'equation de Dirac dans l'alg\`ebre de Pauli\/}
Ann. de la Fond. Louis de Broglie 22(1) 1997\\
{\it Dirac equation in the Clifford algebra of space\/}
in proceedings of ``Clifford algebras and their applications in
mathematical physics'' Aachen 1996, V.Dietrich, K. Habetha, G. Jank eds.
Kluwer /Dordrecht 1998 67--87
%
\bibitem[Dirac 1928]{Dirac-1}
P.A.M. Dirac; {\it The quantum theory of the electron\/}
Proc. Royal Soc. London A117 1928 610
%
\bibitem[Dyson 1972]{Dyson} 
F. Dyson; {\it Missed opportunities\/} 
Bull. Amer. Math. Soc. 78 1972 635--652
%
\bibitem[Fauser 1999a]{Fauser-envalg}
B. Fauser; {\it Enveloping algebra and bi-module structure of space 
Clifford algebra\/} in preparation
%
\bibitem[Fauser et al. 1999b]{Fauser-abla}
B. Fauser, R. Ab{\l}amowic; {\it On the decomposition of Clifford 
algebras with arbitrary bilinear form\/} in proceedings of ``Clifford 
algebras and their applications in mathematical physics'' Ixtapa, 
Mexico 1999, Vol 1, R. Ablamowicz, B. Fauser eds. submitted
%
\bibitem[Fauser et al. 1999c]{Fauser-iso}
B. Fauser, H. Dehnen; {\it Isospin from Spin by Compositenes\/}
in proceedings of ``International Workshop on Lorentz Group, CPT, and
Neutrinos''\/ Zacatecas /Mexico 1999 in press hep-th/9908015
%
\bibitem[Fauser 1999d]{Fauser-hecke}
B. Fauser; {\it Hecke algebra representations within Clifford geometric 
algebras of multivectors\/} J. Phys. A: Math. Gen. 32 (1999) 1919--1936 
%
\bibitem[Fauser 1998]{Fauser-dirac}
B. Fauser; {\it Dirac theory from a field theoretic point of view\/} 
in ``Clifford algebras and their applications in mathematical physics'' 
Aachen 1996, V. Dietrich, K. Habetha, G. Jank eds., Kluwer /Dordrecht 
1998 p. 89--107
%
\bibitem[Fauser 1997]{Fauser-vacua}
B. Fauser; {\it Clifford geometric parameterization of inequivalent vacua\/}
preprint hep-th/9710047 revised at
{\underline{\tt http://kaluza.physik.uni-konstanz.de/\~{}fauser/}}
%
\bibitem[Fock 1929]{Fock} 
V. Fock; {\it Geometrisierung der Diracschen Theorie des Elektrons\/}
Z.Phys. 55 1929 261\\
{\it Sur les \'equations de Dirac dans la th\'eorie de relativit\'e
g\'en\'erale\/} C.R. Acad. Sciences (Paris) 189 1929 25\\
\& D. Ivanenko {\it \"Uber eine m\"ogliche Deutung der relativistischen
Quantentheorie\/} Z.Phys. 54 1929 798\\
{\it G\'eom\'etrie quantique lin\'eaire et d\'eplacement parall\'ele\/}
C.R. Acad. Sciences (Paris) 188 1929 1470
%
\bibitem[Haft 1996]{Haft}
M. Haft; {\it Konjugationen und Diskrete Symmetrien\/} Thesis, LMU Munich
1996
%
\bibitem[Hahn 1994]{Hahn}
A.J. Hahn; {\it Quadratic algebras, Clifford algebras, and arithmetic 
Witt groups\/} Springer /Berlin 1994
%
\bibitem[Hestenes 1995]{Hest-n}
D Hestenes; {\it Real Dirac theory\/} in proceedings of ``The Theory of
the Electron'' Cuautitlan / Mexico 1995, J. Keller, Z. Oziewicz eds. Adv.
in Appl. Clifford Alg. 7(Suppl.) 1997 97--144 and references there
%
\bibitem[Hestenes 1985]{Hest-ulmp} 
D. Hestenes; {\it A unified language for mathematics and physics\/}
in proceedings of ``Clifford algebra and their application in mathematical
physics'' Canterbury /UK. 1985 J.S.R. Chisholm, A.K. Common eds. Kluwer
/Dordrecht 1986 p 1
%
\bibitem[Hestens et al. 1984]{HestSob}
D. Hestenes, G. Sobczyk; {\it Clifford algebra to geometric calculus -- a
unified language for mathematics and physics\/} Reidel /Dordrecht
[1984] 2nd ed. 1992 
%
\bibitem[Hestenes 1967]{Hestenes-iso}
D. Hestenes; {\it Spin and Isospin\/} J. Math. Phys. 8(4) 1967 809
%
\bibitem[Hestenes 1966]{Hest-1}
D. Hestenes; {\it Space time algebra\/} Gordon \& Breach 1966
%
\bibitem[Isham 1995]{Isham}
C. Isham; {\it  Lectures on quantum theory -- mathematical and
structural foundations\/} Imperial College Press/River Edge, NJ 1995
%
\bibitem[Keller 1993]{Keller}
J. Keller; {\it The geometric content of the electron theory, part 1\/}
Adv. in Appl. Clifford Alg. 3(2) 1993 147--200
%
\bibitem[Lounesto 1997]{Lounesto}
P. Lounesto; {\it Clifford algebras and spinors\/} Cambridge Univ. 
Press /Cambridge 1997
%
\bibitem[Madelung 1929]{Madelung}
E. Madelung; {\it Eine \"Ubertragung der Diracschen Theorie des Elektrons
in gewohnte Formen\/}
Z.Phys. 54 1929 303
%
\bibitem[Parra]{Parra-dirac}
J.M. Parra-Serra; {\it Dirac's theory in real geometric formalism:
multivectors versus spinors\/} unpublished, available at 
\underline{\tt http://hermes.ffn.ub.es/\~{}jmparra/}
%
\bibitem[Parra 1989]{Parra-conf}
J.M. Parra-Serra; {\it On Dirac and Darwin-Hestenes equation\/}
in proceedings of ``Clifford algebras and their application in
mathematical physics'' Montpellier 1989, A.Micali, R.Boudet, J.Helmstetter
eds. Kluver/Dordrecht 1990 p 463
%
\bibitem[Pauli 1933]{Pauli-hdb}
W. Pauli; {\it Handbuch der Physik\/} 2.Aufl. Band 24
p 1 Springer /Berlin 1933
%
\bibitem[Rodrigues et al. 1996]{Rodrigues} 
W.A. Rodrigues Jr., Q.A.G. de Souza, J. Vaz Jr., P. Lounesto {\it
Dirac-Hestenes spinor fields in Riemann-Cartan spacetime\/}
Int. J. Theor. Phys. 35 1996 1849--1900  
%
\bibitem[Rylov 1995]{Rylov}
Y.A. Rylov; {\it Dirac equation in terms of hydrodynamical variables\/}
Adv. in Appl. Clifford Alg. 5(1) 1995 1--40
%
\bibitem[Saller 1996]{Saller}
H. Saller; {\it Logik und Quantenstruktur\/} preprint MPI-PhT/1996
MPI-Munich\\
{\it Brevier\/} unpublished book ver. from 1996
%
\bibitem[Sallhofer 1991]{Sallhofer}
H. Sallhofer; {\it Hydrogene in electrodynamics VII\/}
Z. Naturforsch. 49a 1994 1074--1076 and references there
%
\bibitem[Simulik et al. 1998]{Simulik}
V.M. Simulik, I.Yu. Krivsky; {\it Bosonic symmetries of the massless Dirac
equations\/} 
Adv. in Appl. Clifford Alg. 8(1) 1998 69--82\\
{\it Clifford algebra in classical electrodynamical hydrogene atom
models\/} Adv. in Appl. Clifford Alg. 7(1) 1997 25--34
%
\bibitem[Takabayasi 1957]{Takabayasi}
T. Takabayasi; {\it Relativistic hydrodynamics of Dirac matter\/}
Suppl. Proc. Theor. Phys. 4 1957 1--50
%
\bibitem[Yvon 1940]{Yvon}
J. Yvon; {\it Equations de Dirac-Madelung\/} J. Phys. et le Radium 1 1940
18--30
\end{thebibliography}
\end{document}